\documentclass[10pt,conference]{IEEEtran}
\IEEEoverridecommandlockouts
\usepackage{cite}
\usepackage{amsmath,amssymb,amsfonts}
\usepackage{algorithmic}
\usepackage{graphicx}
\usepackage{textcomp}
\usepackage{xcolor}
\usepackage{booktabs}
\usepackage{multirow}
\usepackage{xspace}
\usepackage{caption}
\usepackage{subcaption}
\usepackage{footnote}
\usepackage{url}
\usepackage{tcolorbox}
\usepackage{array}

\def\BibTeX{{\rm B\kern-.05em{\sc i\kern-.025em b}\kern-.08em
    T\kern-.1667em\lower.7ex\hbox{E}\kern-.125emX}}
\begin{document}

\title{Towards Automating Code Review Activities}

\author{
\IEEEauthorblockN{Rosalia Tufano\IEEEauthorrefmark{1}, Luca Pascarella\IEEEauthorrefmark{1}, Michele Tufano\IEEEauthorrefmark{2}, Denys Poshyvanyk\IEEEauthorrefmark{3}, Gabriele Bavota\IEEEauthorrefmark{1}}

\IEEEauthorblockA{\IEEEauthorrefmark{1}\textit{SEART @ Software Institute, Universit\`{a} della Svizzera italiana (USI), Switzerland}}
\IEEEauthorblockA{\IEEEauthorrefmark{2}\textit{Microsoft, USA}}
\IEEEauthorblockA{\IEEEauthorrefmark{3}\textit{SEMERU @ Computer Science Department, William and Mary, USA}}
}

\maketitle

\newcommand{\approach}{\emph{APPROACH}\xspace}
\newcommand{\ie}{\emph{i.e.,}\xspace}
\newcommand{\eg}{\emph{e.g.,}\xspace}
\newcommand{\etc}{etc.\xspace}
\newcommand{\etal}{\emph{et~al.}\xspace}
\newcommand{\secref}[1]{Section~\ref{#1}\xspace}
\newcommand{\figref}[1]{Fig.~\ref{#1}\xspace}
\newcommand{\listref}[1]{Listing~\ref{#1}\xspace}
\newcommand{\tabref}[1]{Table~\ref{#1}\xspace}
\newcommand{\tool}[1]{{\sc #1}\xspace}

\newcommand{\codePairs}{17,194\xspace}

\newboolean{showcomments}

\setboolean{showcomments}{true}

\ifthenelse{\boolean{showcomments}}
  {\newcommand{\nb}[2]{
    \fbox{\bfseries\sffamily\scriptsize#1}
    {\sf\small$\blacktriangleright$\textit{#2}$\blacktriangleleft$}
   }
  }
  {\newcommand{\nb}[2]{}
  }

\newcommand\MICHELE[1]{\textcolor{red}{\nb{MICHELE}{#1}}}
\newcommand\ROSALIA[1]{\textcolor{red}{\nb{ROSALIA}{#1}}}
\newcommand\LUCA[1]{\textcolor{red}{\nb{LUCA}{#1}}}
\newcommand\DENYS[1]{\textcolor{red}{\nb{DENYS}{#1}}}
\newcommand\GABRIELE[1]{\textcolor{red}{\nb{GABRIELE}{#1}}}

\newcommand\revised[1]{\textcolor{black}{#1}}

\begin{abstract}
Code reviews are popular in both industrial and open source projects. The benefits of code reviews are widely recognized and include better code quality and lower likelihood of introducing bugs. However, since code review is a manual activity it comes at the cost of spending developers' time on reviewing their teammates' code.

Our goal is to make the first step towards partially automating the code review process, thus, possibly reducing the manual costs associated with it. We focus on both the \emph{contributor} and the \emph{reviewer} sides of the process, by training two different Deep Learning architectures. The first one learns code changes performed by developers during real code review activities, thus providing the \emph{contributor} with a revised version of her code implementing code transformations usually recommended during code review before the code is even submitted for review. The second one automatically provides the \emph{reviewer} commenting on a submitted code with the revised code implementing her comments expressed in natural language. 

The empirical evaluation of the two models shows that, on the contributor side, the trained model succeeds in replicating the code transformations applied during code reviews in up to 16\% of cases. On the reviewer side, the model can correctly implement a comment provided in natural language in up to 31\% of cases. \revised{While these results are encouraging, more research is needed to make these models usable by developers}.
\end{abstract}


\section{Introduction} \label{sec:intro}
Code Review is the process of analyzing source code written by a teammate to judge whether it is of sufficient quality to be integrated into the main code trunk. Recent studies provided evidence that reviewed code has lower chances of being buggy \cite{McIntosh:msr2014,morales2015saner,Bavota:icsme2015} and exhibits higher internal quality \cite{Bavota:icsme2015}, likely being easier to comprehend and maintain. Given these benefits, code reviews are widely adopted both in industrial and open source projects with the goal of finding defects, improving code quality, and identifying alternative solutions. 

The benefits brought by code reviews do not come for free. Indeed, code reviews add additional expenses to the standard development costs due to the allocation of one or more reviewers having the responsibility of verifying the correctness, quality, and soundness of newly developed code. Bosu and Carver report that developers spend, on average, more than six hours per week reviewing code \cite{Bosu:2013}. This is not surprising considering the high number of code changes reviewed in some projects: Rigby and Bird \cite{Rigby:fse2013} show that industrial projects, such as Microsoft Bing, can undergo thousands of code reviews per month ($\sim$3k in the case of Bing). Also, as highlighted by Czerwonka \etal \cite{Czerwonka:icse2015}, the effort spent in code review does not only represent a cost in terms of time, but also pushes developers to switch context from their tasks. 

Our long-term goal is to reduce the cost of code reviewing by (partially) automating this time-consuming process. Indeed, we believe that several code review activities can be automated, such as, catching bugs, improving adherence to the project's coding style, and refactoring suboptimal design decisions. The final goal is not to replace developers during code reviews but work with them in tandem by automatically solving (or suggesting) code quality issues that developers would manually catch and fix in their final checks. A complete automation, besides likely not being realistic, would also dismiss one of the benefits of code review: the sharing of knowledge among developers.

In this paper, we make a first step in this direction by using Deep Learning (DL) models to partially automate specific code review tasks. First, from the perspective of the \emph{contributor} (\ie the developer submitting the code for review), we train a transformer model \cite{vaswani2017attention}  to ``translate'' the code submitted for review into a version implementing code changes that a reviewer is likely to suggest. In other words, we learn code changes recommended by reviewers during review activities and we try to automatically implement them on the code submitted for review. This could give a fast and preliminary feedback to the contributor as soon as she submits the code. This model has been trained on \codePairs code pairs of $C_s$ $\rightarrow$ $C_r$ where $C_s$ is the code submitted for review and $C_r$ is the code implementing a specific comment provided by the reviewer. 

Once trained, the model can take as input a previously unseen code and recommend code changes as a reviewer would do. The used architecture is a classic encoder-decoder model with one encoder taking the submitted code as input and one decoder generating the revised source code.

Second, from the perspective of the \emph{reviewer}, given the code under review ($C_s$) we want to provide the ability to automatically generate the code $C_r$ implementing on $C_s$ a specific recommendation expressed in natural language ($R_{nl}$) by the reviewer. This would allow (i) the reviewer to automatically attach to her natural language comment a preview of how the code would look like by implementing her recommendation, and (ii) the contributor to have a better understanding of what the reviewer is recommending. For such a task, we adapt the previous architecture to use two encoders and one decoder. 

The two encoders take as input $C_s$ and $R_{nl}$, respectively, while the decoder is still in charge of generating $C_r$. The model has been trained with \codePairs triplets $\langle$$C_s$, $R_{nl}$$\rangle$ $\rightarrow$ $C_r$.\newpage

Note that the two tackled problems are two sides of the same coin: In the first scenario, $C_r$ is generated without any input provided by the reviewer, thus allowing the usage of the model even before submitting the code for review. In the second scenario, $C_r$ is generated with the specific goal of implementing a comment provided by the reviewer, thus when the code review process has already started.

We quantitatively and qualitatively evaluate the predictions provided by the two approaches. For the quantitative analysis, we assessed the ability of the models in modifying the code submitted for review exactly as done by developers during real code review activities. This means that we compare, for the same code submitted for review, the output of the manual code review process and of the models (both in the scenario where a natural language comment is provided or not as input). The qualitative analysis focuses instead on characterizing successful and unsuccessful predictions made by the two models, to better understand their limitations.

The achieved results indicate that, in the \emph{contributor} scenario (1-encoder model), the model can correctly recommend a change as a reviewer would do in 3\% to 16\% of cases, depending on the number of candidate recommendations it is allowed to generate. When also having available a reviewer comment in natural language (\ie \emph{reviewer} scenario, 2-encoder model), the performances of the approach are boosted, with the generated code that correctly implements the reviewer's comment in 12\% to 31\% of cases. \revised{These preliminary results can pave the way to novel research in the area of automating code review.}
\section{Using Transformers to Automate Code Review} \label{sec:approach}
\begin{figure*}
	\centering
	\includegraphics[width=0.95\linewidth]{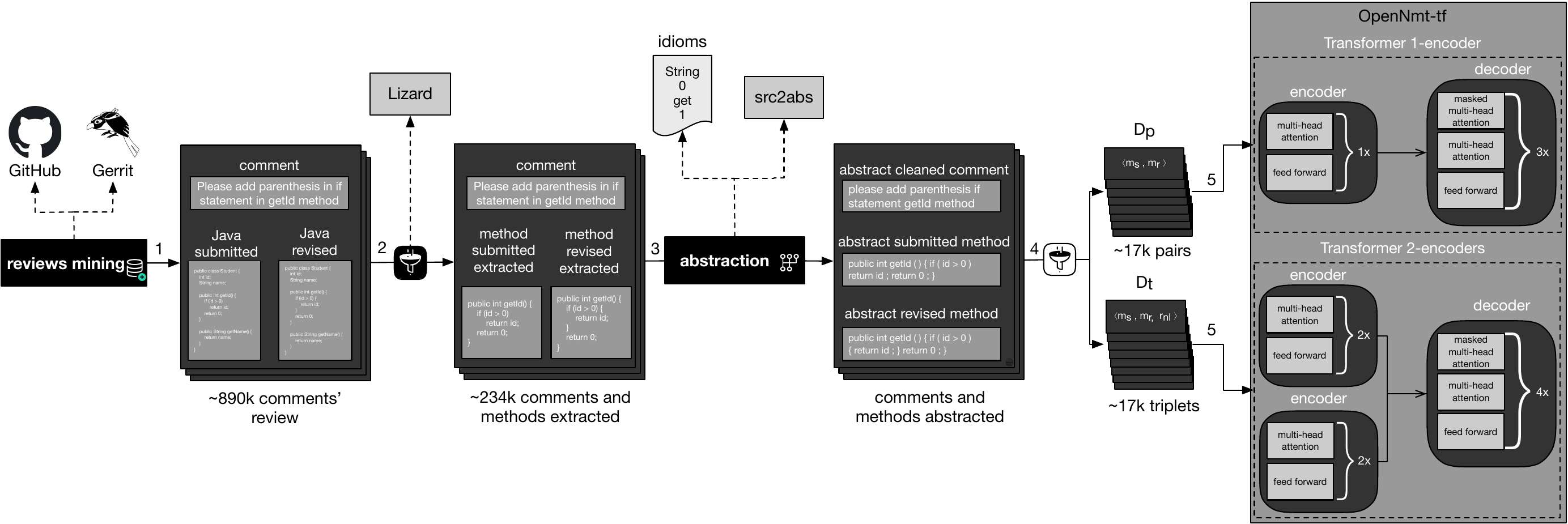}
	\caption{Approach overview.}
	\label{fig:approach}
\end{figure*}

\figref{fig:approach} shows the basic steps of our approach. In a nutshell, we start by mining code reviews from Java projects hosted on GitHub \cite{GitHub} and/or using Gerrit \cite{Gerrit} as code review platform (Step 1 in \figref{fig:approach}). Given a code submitted by a contributor for review, we parse it to extract the list of methods it contains. Indeed, in this first work on automating code reviews, we decided to focus on small and well-defined code units represented by methods. We identify all submitted methods $m_s$. Then, we collect reviewer's comments made on each $m_s$ by exploiting information available in both GitHub and Gerrit linking a reviewer comment to a specific source code line. We refer to each of those comments as $r_{nl}$ (\ie a natural language recommendation made by a reviewer). In such a phase, a set of filters is applied to automatically discard comments unlikely to recommend and results in code changes (\eg ``thank you'', ``well done'', \etc) (2).

If the contributor decides to address (some of) the received $r_{nl}$, this will result in a revised version of $m_s$ addressing the received comments. We refer to such a version as $m_r$. Both $m_s$ and $m_r$ are abstracted to reduce the vocabulary size and make them more suitable for DL  \cite{Tufano:icse2019,Tufano:tosem2019,Watson:icse2020} (3). To increase the likelihood that $m_r$ actually implements in $m_s$ a specific reviewer's comment, we only consider $m_s$ that received a single comment in a review round. Thus, if a revised version of $m_s$ is submitted, we can conjecture that it implements a single comment received by a reviewer (4). 

Such a process results in a dataset of Reviewed Commented Code Triplets (RCCTs) in the form $\langle$$m_s$, $r_{nl}$$\rangle$ $\rightarrow$ $m_r$. Such a dataset is used to train a transformer architecture using two encoders (one processing $m_s$ and one $r_{nl}$) and one decoder (generating $m_r$). Such a model is able, given a Java method ($m_s$) and a reviewer comment about it ($r_{nl}$) to generate a revision of $m_s$ implementing $r_{nl}$ (\ie $m_r$) (5). 

Starting from this dataset, we also generate a dataset of code pairs $m_s$ $\rightarrow$ $m_r$ obtained by removing $r_{nl}$ from each of the previous dataset triplets. This dataset has been used to train a second transformers-based model having one encoder (processing $m_s$) and one decoder (generating $m_r$). Once trained, this model can take as input a previously unseen Java method ($m_s$) and recommend a revised version of it ($m_r$) that would likely result from a review round. Since no input is required from the reviewer in this model, it can be used by the contributor to double check her implementation before submitting it for review.

Next, we describe the different steps behind our approach. 

\subsection{Mining Code Review Data}
\label{sec:mining}
We built two crawlers for mining from Gerrit and GitHub code review data. Before moving to the technical details, it is important to clarify what the goal of this mining process is. Once a code contribution (\ie changes impacting a set of existing code files or resulting in new files) is submitted for review, it can be a subject to several review rounds. Let us assume that $C_s$ is the set of code files submitted for review, since subject to code changes. A set of reviewer comments \{$r_{nl}$\} can be made on $C_s$ and, if some/all of them are addressed, this will result in a revised version of the code $C_{r1}$. This is what we call a ``review round'', and can be represented by the triplet $\langle$$C_s$ , \{$r_{nl}$\}$\rangle$ $\rightarrow$ $C_{r1}$. The completion of a review round does not imply the end of the review process. Indeed, it is possible that additional comments are made by the reviewers on $C_{r1}$ and that those comments are addressed. This could result, for example, in a second triplet $\langle$$C_{r1}$ , \{$r_{nl}$\}$\rangle$ $\rightarrow$ $C_{r2}$. The goal of our mining is to collect all triplets output of the code review rounds performed in Gerrit and in GitHub. 

To this aim, we developed two miner tools tailored for systematically querying Gerrit and GitHub public APIs. The double implementation is required, because despite the fact that both platforms provide a similar support for code review, the public APIs used to retrieve data differ. Gerrit does not offer an API to retrieve all the review requests for a given project, but it is possible to retrieve them for an entire Gerrit installation (\ie an installation can host several projects, such as all Android-related projects). Starting from this information, we collect all ``review rounds'', and finally, we reorganized the retrieved data by associating the own set of reviews to each project. Overall, we mined six Gerrit installations, for a total of 6,388 projects.

GitHub instead offers an API to collect a list of ids of all review requests per project. In this case, we mined a set of 2,566 GitHub Java repositories having at least 50 PRs obtained by querying the GitHub APIs.

The output of this process is represented, for each review round, by (i) the set of code files submitted for review, (ii) the comments received on this code files with information about the specific impacted lines (character-level information is available for Gerrit), and (iii) the revised code files submitted in response to the received comments. 

\subsection{Data Preprocessing}
\label{sec:processing}
After having collected the data from Gerrit and GitHub, we start its preprocessing, with the goal of building the two datasets of triplets ($\langle$$m_s$, $r_{nl}$$\rangle$ $\rightarrow$ $m_r$) and pairs ($m_s$ $\rightarrow$ $m_r$) previously mentioned.

\subsubsection{Methods Extraction and Abstraction}
We start by parsing the Java files involved in the review process (both the ones submitted for review and the ones implementing the code review comments) using the Lizard \cite{Lizard} Python library. The goal of the parsing is to extract the methods from all the files. Indeed, as said, we experiment with the DL models at method-level granularity, as also done in previous work \cite{Tufano:icse2019,Tufano:tosem2019,Watson:icse2020}. After this step, for each mined review round, we have the list of Java methods submitted for review, the reviewers' comments, and the revised list of methods resubmitted by the author to address (some of) the received comments. 

\begin{figure}
	\centering
	\includegraphics[width=0.75\linewidth]{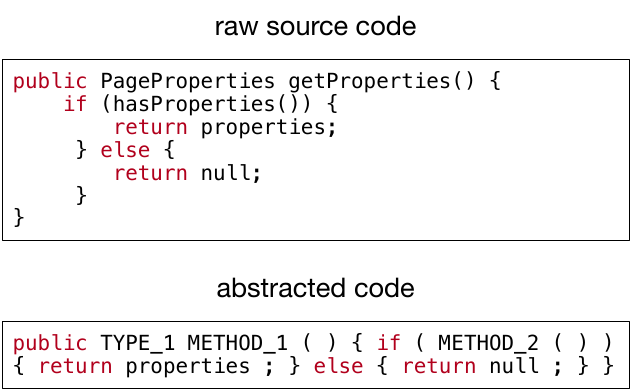}
	\caption{Example of abstraction.}
	\label{fig:abstraction}
	\vspace{-0.1cm}
\end{figure}

Then, we adopt the abstraction process described in the work by Tufano \etal \cite{Tufano:icse2019} to obtain a vocabulary-limited yet expressive representation of the source code. Recent work on generating assert statements using DL \cite{Watson:icse2020} showed that the performance of sequence-to-sequence models on code is substantially better when the code is abstracted with the procedure presented in \cite{Tufano:icse2019} and implemented in the \texttt{src2abs} tool \cite{src2abs}. Triplets for which a parsing error occur during the abstraction process on the $m_s$ or on the $m_r$ methods are removed from the dataset. \figref{fig:abstraction} shows an example of abstraction procedure we perform. The top part represents the raw source code. \texttt{src2abs} uses a Java lexer and a parser to represent each method as a stream of tokens, in which Java keywords and punctuation symbols are preserved and the role of each identifier (\eg whether it represents a variable, method, \etc) as well as the type of a literal is discerned. 

IDs are assigned to identifiers and literals by considering their position in the method to abstract: The first variable name found will be assigned the ID of VAR\_1, likewise the second variable name will receive the ID of VAR\_2. This process continues for all identifiers as well as for the literals (\eg STRING\_X, INT\_X, FLOAT\_X). Since some identifiers and literals appear very often in the code (\eg variables \texttt{i}, \texttt{j}, literals \texttt{0}, \texttt{1}, method names such as \texttt{size}), those are treated as ``idioms'' and are not abstracted. We construct our list of idioms by looking for the 300 most frequent identifiers and literals in the extracted methods (list available in our replication package \cite{replication}). The bottom part of \figref{fig:abstraction} shows the abstracted version of the source code. Note that during the abstraction code comments are removed. \texttt{src2abs} is particularly well suited for the abstraction in our context, since it implements a ``pair abstraction mode'', in which a pair of methods can be provided (in our case, $m_s$ and $m_r$) and the same literals/identifiers in the two methods will be abstracted using the same IDs. As output of the abstract process, \texttt{src2abs} does also provide an abstraction map $M$ linking the abstracted token to the raw token (\eg mapping $VAR\_1$ to $sum$). This allows to go back to the raw source code from the abstracted one \cite{Tufano:icse2019}.

\subsubsection{Linking and Abstracting Reviewer Comments}
Each collected reviewer comment is associated with the specific set of code lines it refers to. This holds both for Gerrit and GitHub. 

Using this information, we can link each comment to the specific method (if any) it refers to: Given $l_s$ and $l_e$ the start and the end line a given comment refers to, we link it to a method $m_i$ if both $l_s$ and $l_e$ fall within $m_i$'s body, signature, or annotations (\eg \texttt{@Override}). If a comment cannot be linked to any method (\eg it refers to an \texttt{import} statement) it is discarded from our dataset, since useless for our scope. 

After having linked comments to methods for each review round, we are in the situation in which we have, for each review round, a set of triplets $\langle$$m_s$, $m_r$, and \{$r_{nl}$\}$\rangle$, where $m_s$ and $m_r$ represent the same abstracted method before and after the review round, and \{$r_{nl}$\} is a set of comments $m_s$ received in this round. At this point, we also abstract all code components mentioned in any comment in \{$r_{nl}$\} using the abstraction map obtained during the abstraction of $m_s$ and $m_r$. Thus, assuming that the comment mentions ``\emph{change the type of sum to double}'' and that the variable \texttt{sum} has been abstracted to $VAR\_1$, the comment is transformed into ``\emph{change the type of $VAR\_1$ to double}''. On top of this, any camel case identifier that is not matched in the abstraction map but that it is present in the comment, is replaced by the special token \_CODE\_. Such a process ensures consistency in (i) the representation of the code and the comment that will be provided as input to the 2-encoder model, and (ii) the representation of similar comments talking about different \_CODE\_ elements. 

\subsubsection{Filtering Out Noisy Comments}
Through a manual inspection of the code review data we collected, we noticed that a non-negligible percentage of code comments we were collecting, while linked to source code lines, were unlikely to result in code changes and, thus, irrelevant for our study. For example, if two reviewers commented on the same method, one saying ``\emph{looks good to me}'' and the other one asking for a change ``\emph{please make this method static}'', it is clear that any revised version of the method submitted afterwards by the contributor would be the result of implementing the second comment rather than the first one. With the goal of minimizing the amount of noisy comments (\ie comments unlikely to result in code changes) provided to our model, we devised an approach to automatically classify a comment as \emph{likely to lead to code changes} (from now on simply \emph{relevant}) or \emph{unlikely to lead to code changes} (\emph{irrelevant}). 

We started by creating a dataset of comments manually labeled as \emph{relevant} or \emph{irrelevant}. To this aim, we randomly selected from our dataset a set of 1,875 comments and related methods $m_s$. These comments come from 500 reviews performed on Gerrit and 500 performed on GitHub. On our dataset (that we will detail later), such a sample guarantees a significance interval (margin of error) of $\pm3\%$ with a confidence level of 99\% \cite{Rosner2011}. The comments have then been loaded in a web-app we developed to support the manual analysis process, that was performed by three of the authors. The web-app assigned each comment to two evaluators and, in case of conflict (\ie one evaluator said that the comment was relevant and one that was irrelevant) the comment was assigned to a third evaluator, that solved the conflict through majority voting. 

Conflicts arose for 21\% of the analyzed comments. Examples of comments labeled as irrelevant include simple and obvious cases such as ``Thanks!'' and ``Nice'', but also more tricky instances difficult to automatically identify (\eg ``At least here it is clear that the equals method of the implementors of TreeNode is important'').

The final labeled dataset consists of 1,875 comments, of which 1,676 have been labeled as \emph{relevant} and 199 as \emph{irrelevant}. We tried to use a simple Machine Learning (ML)-based approach to automatically classify a given comment as relevant or not. We experimented with many different variants of ML-based techniques for this task. As predictor variables (\ie features) of each comment, we considered $n$-grams extracted from them, with $n \in \{1, 2, 3\}$. Thus, we consider single words as well as short sequences of words (2-grams and 3-grams) in the comment. Before extracting the $n$-grams, the comment text is preprocessed to remove punctuation symbols and put all text to lower case. In addition to this, only when extracting 1-grams, English stopwords \cite{Stopword} are removed and the Porter stemmer \cite{Porter:program1980} is applied to reduce all words to their root. These two steps are not performed in the case of 2- and 3-grams, since they could break the ``meaning'' of the extracted n-gram (\eg from a comment ``\emph{if condition should be inverted}'' we extract the 2-gram ``if condition''; by removing stopwords, the if would be removed, breaking the 2-gram). Finally, in all comments we abstract the mentioned source code components as previously explained. 

After having extracted the features, we trained the {\em Weka} \cite{WEKA} implementation of three different models, \ie the Random Forest, J48, and Bayesian network \cite{Breiman:2001} to classify our comments. We performed a 10-fold cross validation to assess the performance of the models. Since our dataset is substantially unbalanced (89\% of the comments are relevant), we re-balanced our training sets in each of the 10-fold iterations using SMOTE \cite{chawla2002smote}, an oversampling method which creates synthetic samples from the minor class. We experimented each algorithm both with and without SMOTE. Also, considering the high number of features we extracted, we perform an {\em information gain} feature selection process \cite{Mitc1997a} aimed at removing all features that do not contribute to the information available for the prediction of the comment type. This procedure consists of computing the information gain of each predictor variable. This value ranges between 0 (\ie the predictor variable does not carry any useful information for the prediction) to 1 (maximum information). We remove all features having an information gain lower than 0.01.

We analyze the results with a specific focus on the precision of the approaches when classifying a comment as \emph{relevant}. Indeed, what we really care about is that when the model classifies a comment as relevant, it is actually relevant and will not represent noise for the DL model. The achieved results reported the Random Forest classifier using the SMOTE filter as the best model, with a precision of 91.6\% (meaning, that $\sim$92 out of 100 comments classified as \emph{relevant} are actually relevant). While such a result may look good, it is worth noting that 89\% of the comments in our dataset are \emph{relevant}. 

This means that a constant classifier always answering ``relevant'' would achieve a 89\% precision. Thus, we experimented with a different and simpler approach. We split the dataset of 1,875 comments into two parts, representing 70\% and 30\% of the dataset. Then, one of the authors tried to define simple keyword-based heuristics with the goal of maximizing the precision in classifying relevant comments on the 70\% subset. Through a trial-and-error process he defined a set of heuristics that we provide as part of our replication package \cite{replication}. In short, these heuristics aim at removing: (i) useless 1-word comments (\eg ``nice''), (ii) requests to change formatting with no impact on code (\eg ``please fix indentation''), (iii) thank you/approval messages (\eg ``looks good to me''), (iv) requests to add test code, that will not result in changes to the code under review (\eg ``please add tests for this method''), (v) requests for clarification (\eg ``please explain''), (vi) references to a previous comment that cannot be identified (\eg ``same as before''), and (vii) requests to add comments, that we ignore in our study (\eg ``add to Javadoc''). Once there was no more room for improvement on the 70\% subset, the set of defined heuristics has been tested on the 30\% dataset, achieving precision of 93.4\% in classifying relevant comments. On the same 30\% test set, the running of a Random Forest trained on the 70\% dataset achieved precision of 92.1\%. Given these results, we decided to use the set of defined heuristics as one of the filtering steps in our approach when preparing the dataset for training our models. Basically, these heuristics remove from the triplets $\langle$$m_s$, $m_r$, \{$r_{nl}$\}$\rangle$ comments in \{$r_{nl}$\} that are unlikely to have triggered the code changes that transformed $m_s$ in $m_r$. 

\subsection{Automating Code Review}
\label{sec:learning}

\subsubsection{Dataset Preparation}
Starting from the collected triplets, our goal is to build two datasets for the training/test of 1- and 2-encoder model. First, we removed from all triplets the comments classified as noisy. Then, we built the dataset for the 2-decoder model since the other one can be easily obtained from it. We apply a set of filtering steps to obtain triplets $\langle$$m_s$, $m_r$, \{$r_{nl}$\}$\rangle$ in which: \smallskip

\emph{\{$r_{nl}$\} does not contain any comment posted by the contributor}. We are interested only in reviewers' comments. Thus, author's comments are removed, leaving \revised{231,439} valid triplets.

\emph{\{$r_{nl}$\} does not contain any comment linked to lines in the related method representing  code comments}. Such $r_{nl}$ are removed from each \{$r_{nl}$\} before the abstraction process since, as previously explained, the abstraction removes comments. 

\emph{$m_s$ and $m_r$, after the abstraction, must be different}. If $m_s$ and $m_r$ are not different, we can remove the triplet, since this means that no code change has been implemented as result of the reviewer's comments. Thus, there is nothing to learn for our models. Such a scenario can happen in the case in which the change is applied to code indentation, code comments, \etc

\emph{$m_s$ and $m_r$ have a reasonable length that can be handled through NMT models}. The variability in sentences length can affect training and performance of NMT models even when techniques such as bucketing and padding are employed. 

Thus, we exclude all triplets having $m_s$ or $m_r$ longer than 100 tokens after abstraction. Such a filtering step has been performed in previous work \cite{Tufano:icse2019,Tufano:tosem2019,Watson:icse2020}, \revised{and it is responsible for the removal of 148,539 triplets from our dataset.}

\emph{$m_r$ does not introduce identifiers or literals that were not present in $m_s$}. If $m_r$ introduces, for example, a new variable \texttt{VAR\_3} that was not present in $m_s$, in a real usage scenario it would not be possible for the model to generate the concrete raw source code for $m_r$, since it could not guess what the actual value for \texttt{VAR\_3} should be. Thus, the model would be useless to developers.  Such a limitation is due to the abstraction process that, however, has the advantage of limiting the vocabulary size and of helping the model learning \cite{Watson:icse2020}. However, the presence of idioms allows to retain in our dataset triplets that otherwise would be discarded because of the inability to synthesize new identifiers/literals in $m_r$.

\emph{\{$r_{nl}$\} is a singleton, meaning that a single comment has been provided by a reviewer on $m_s$}. All triplets containing more than one comment in \{$r_{nl}$\} have been removed, since in those cases we cannot know what was the comment that triggered the transformation of $m_s$ in $m_r$.\smallskip

The remaining triplets are thus in the form $\langle$$m_s$, $m_r$, $r_{nl}$$\rangle$. We preprocess the $r_{nl}$ comment to remove from it stopwords \cite{Stopword}, and links identified through regular expressions (\eg links pointing to online examples). Then, we clean the comment from superfluous punctuation like an ending question mark. At the end, we transform all comment words that are not code IDs in lower case, \eg the comment ``\emph{Could we use String instead of Text?}'' is transformed into ``\emph{String instead $TYPE\_1$}''. Finally, we remove duplicates from the dataset.

After this process, the remaining 17,194 triplets represent what we call the $D_t$ dataset, used for training/evaluating the 2-encoder model. By removing from each triplet the $r_{nl}$ comment, we obtain the $D_p$ dataset, that is instead used to train and evaluate the 1-encoder model. Besides this difference in the two datasets, the code $m_s$ in $D_t$  includes two special tokens $<$START$>$ and $<$END$>$ which mark the part of the code interested by the reviewer's comment $r_{nl}$. These tokens are removed in the $D_p$ dataset, since the 1-encoder model should be used in a scenario in which no comments have been provided by the reviewer yet. Both datasets have been split into training (80\%), evaluation (10\%) and test (10\%) sets.

\subsubsection{Scenario 1: Recommending Changes (1-encoder)}
The 1-encoder model is meant to help the developer anticipating the changes a reviewer might suggest on the submitted code. Therefore, we want to learn how to automatically generate $m_r$ given $m_s$. For this task we use a transformer model \cite{vaswani2017attention}. The transformer model consists of an encoder and decoder, which takes as input a sequence of tokens and generates another sequence as output, but it only relies on the attention-mechanism, without implying any recurrent networks. Both encoder and decoder consist of multiple layers each of which is composed of Multi-Head Attention and Feed Forward layers. 

In this first scenario we train a transformer model with one encoder that will take as input the sequence $m_s$ and one decoder that will generate one or multiple suggestions for $m_r$. 

\subsubsection{Scenario 2: Implementing Changes Recommended by the Reviewer (2-encoder)}
The idea for the second scenario is to automatically implement a reviewer recommendation expressed in natural language in order to produce a practical example of what the reviewer wants. Therefore, given $m_s$ and $r_{nl}$ we want to automatically generate the sequence $m_r$. Also for this task we  train a transformer model, but using two encoders and one decoder. The two encoders take as input the sequences $m_s$ and $r_{nl}$, respectively, while the decoder generates one or multiple suggestions for $m_r$. To implement both models we used the Python library OpenNmt-tf \cite{klein2018opennmt,OpenNmt-tf}.

\subsubsection{Hyperparameter Search}
\label{sec:hyperpar}
For both models, in order to find the best configurations, we performed hyper-parameter search by adopting a Bayesian Optimization strategy \cite{snoek2012practical,hutter2011sequential}. We created the space of possible configurations selecting the 10 hyper-parameters reported in \tabref{tab:hyperparameters} and choosing for each one an interval of possible values by looking at the DL literature. Given the large size of the domain space, to explore it, we chose the Tree Parzen Estimator (TPE) \cite{NIPS2011_4443,bergstra2013making} as optimization algorithm with the maximum number of trials equals to 40. This means that 40 different configuration of hyper-parameters have been tested for each model. Each configuration has been trained for a maximum of 50k steps using the number of perfect predictions on the evaluation set as optimization metric. This means that the best configuration output of this process is the one for which the model is able to generate the highest number of $m_r$ strings that are identical to the ones written by developers. To support this process, we used the Hyperopt Python library \cite{bergstra2013hyperopt,Hyperopt}.

\begin{table}
\centering
\scriptsize
\begin{tabular}{llrr} 
\toprule
\textbf{Hyperparameter} & \textbf{Possible values} & \textbf{1-encoder} & \textbf{2-encoder} \\
\midrule
\textbf{Embedding size} & $[128, 256, 512, 1024, 2048]$ & 256 & 512 \\
\textbf{Encoder layers} & $[1, 2, 3, 4]$ & 1 & 2\\
\textbf{Decoder layers} & $[1, 2, 3, 4]$ & 2 & 4\\
\textbf{Number of units} & $[128, 256, 512, 1024, 2048]$ & 256 & 512\\
\textbf{Ffn dimension} & $[128, 256, 512, 1024, 2048]$ & 256 & 512\\
\textbf{Number of heads} & $[2, 4, 8]$ & 8 & 4\\
\textbf{Learning rate} & $(0.0, 1.0)$ & 0.5132 & 0.3370\\
\textbf{Dropout} & $(0.0, 0.4)$ & 0.2798 & 0.1168\\
\textbf{Attention dropout} & $(0.0, 0.4)$ & 0.1873 & 0.1794\\
\textbf{Ffn dropout} & $(0.0, 0.4)$ & 0.2134 & 0.2809\\
\bottomrule
\end{tabular}
\caption{Hyperparameters and the best configuration}
\label{tab:hyperparameters}
\vspace{-0.2cm}
\end{table}

\subsubsection{Generating Multiple Solutions via Beam Search} Once the best configuration of each model has been selected, we evaluate it on the unseen samples of the test set. With the idea that the outputs generated by the models must be suggestions for developers/reviewers, we adopt a Beam Search decoding strategy \cite{bahdanau2014neural,boulanger2013audio,graves2012sequence,raychev2014code} to generate multiple hypotheses for a given input. An output sequence is generated by adding the most likely token given the previous ones step by step. Beam search, instead of considering only the sequence of tokens with the best probability, considers the top-$k$ more probable hypotheses, where $k$ is known as the beam size. Thus, beam search builds $k$ sequences simultaneously. At each timestep, it explores the space of possible hypotheses, consisting of the sequences obtainable by adding a single token to the previous $k$ partial sequences. The process ends when the $k$ sequences are completed. We experiment with beam sizes $k = 1, 3, 5, 10$.
\section{Study Design} \label{sec:design}
The {\em goal} of this study is to empirically assess whether NMT can be used to partially automate code review activities. The {\em context} consists of the $D_p$ and $D_t$ datasets (\secref{sec:approach}).

The study aims at tackling the following research questions:

\begin{itemize}
\item RQ$_1$: \emph{To what extent is NMT a viable approach to automatically recommend to developers code changes as reviewers would do?} This RQ focuses on the ``contributor perspective'' described in the introduction. We evaluate the ability of an NMT model to automatically suggest code changes for a submitted code contribution as reviewers would do. We do not focus on generating the natural language comment explaining the code changes that a reviewer would require, but on providing to the developer submitting the code $C_s$ a revised version of it ($C_r$) that implements changes that will be likely required in the review process. We employ the $D_p$ dataset in the context of RQ$_1$.

\item RQ$_2$: \emph{To what extent is NMT a viable approach to automatically implement changes recommended by reviewers?} The second RQ focuses on the previously described ``reviewer perspective'', and assesses the ability of the NMT model to automatically implement in a submitted code $C_s$ a recommendation provided by a reviewer and expressed in natural language ($R_{nl}$), obtaining the revised code $C_r$. We employ the $D_t$ dataset in the context of RQ$_2$.
\end{itemize}

\subsection{Data Collection and Analysis}
To answer RQ$_1$ we run the best configuration of the 1-encoder model obtained after hyperparameter tuning (\secref{sec:hyperpar}) on the test set of the $D_p$ dataset, and we perform an inference of the model using beam search \cite{Raychev:2014:CCS:2594291.2594321}. 

Given the code predicted by the NMT model, we consider a prediction as correct if it is identical to the code  manually written by a developer after a review round (we refer to these cases as ``perfect predictions''). Since we experiment with different beam sizes, we check whether a perfect prediction exists within the $k$ generated solutions. We report the raw counts and percentages associated with perfect predictions for each beam size. 

Besides reporting the perfect predictions, we also compute the BLEU-4 score \cite{Papineni:2002} of all  predictions. The BLEU score is a metric used for assessing the quality of text automatically generated in the context of a NMT task \cite{Papineni:2002}. It takes values between 0\% and 100\%, where 100\% indicates a perfect prediction, meaning that the predicted text is identical to the reference one. We use the BLEU-4 variant, computed by considering the 4-grams in the generated text and previously used in other software engineering papers (\eg \cite{Watson:icse2020,Tufano:tosem2019}). 

Also, to assess the effort needed by developers to convert a prediction generated by the model into the reference (correct) code, we compute the Levenshtein distance \cite{levenshtein1966} at token-level. This is the minimum number of token edits (insertions, deletions or substitutions) needed to convert the predicted code into the reference one. 

Since such a measure is not normalized, it is difficult to interpret. For this reason, we normalize such a value by dividing it by the number of tokens in the longest sequence among the predicted and the reference code.

Finally, we complement our quantitative data with a qualitative analysis aimed at reporting (i) examples of perfect predictions, categorized based on the type of code change that the model automatically implemented; and (ii) non-perfect predictions, to understand whether they still can be valuable for developers. Concerning the first point, two of the authors manually analyzed all 271 perfect predictions independently, and categorized them by assigning to each prediction a label describing the change automatically injected by the model. Conflicts, that arose in 11\% of cases, have been solved through an open discussion. We present the obtained taxonomy as an output of this analysis. As for the second point, we use the BLEU score ranges 0-24, 25-49, 50-74 and 75-99 to split the imperfect predictions. Then, we randomly selected 25 instances from each set and the first two authors manually evaluated them to determine if the recommended code change is still meaningful while being different to the reference one. Also in this case, conflicts  (\ie cases in which the two authors consistently disagreed) that arose in 9\% of cases were solved through open discussion. 

To answer RQ$_2$ we run the exact same analysis described for RQ$_1$, but by using the $D_t$ dataset. The main differences are related to the performed qualitative analysis. When evaluating the perfect predictions, we decided to focus on the perfect predictions obtained by the 2-encoder model but not by the 1-encoder model. Indeed, those are most likely the cases in which the comment provided as input played a role in the prediction. For those 300 instances, the two authors labeled the reviewers' comments to assign a label expressing the type of code change required by the reviewer. In other words, while in RQ$_1$ the goal was on categorizing the type of code change implemented by the model, here the focus is on the type of change requested by the reviewer in the comment. The goal is to identify categories of code comments that help the model in correctly implementing the required code change. In this case, conflicts arose for 8\% of cases. The second difference, still related to the qualitative analysis, is represented by analyzed failure cases: Here the goal was to check whether the change implemented by the model, while different from the one manually implemented by the developer, was still a meaningful implementation of the change requested by the reviewer (conflicts in 2\% of the analyzed instances). 
\section{Results Discussion} \label{sec:results}
\begin{table*}
\centering
\begin{tabular}{ccccccccccc}
\toprule
\textbf{Beam} & \multicolumn{2}{c}{\textbf{Perfect Predictions}} & & \multicolumn{3}{c}{\textbf{BLEU-4}} & & \multicolumn{3}{c}{\textbf{Levenshtein distance}}\\\cline{2-3} \cline{5-7} \cline{9-11}
\textbf{Size} & \textbf{\#} & \textbf{\%} & & \textbf{mean} & \textbf{median} & \textbf{st. dev.} & & \textbf{mean} & \textbf{median} & \textbf{st. dev.}\\\hline
\midrule
\multicolumn{11}{c}{ {\bf 1-encoder} }\\
\midrule
\textbf{1} & 50 & 2.91\% & & 0.7706 & 0.8315 & 0.1929 & & 0.2383 & 0.2000 & 0.1670\\
\textbf{3} & 156 & 9.07\% & & 0.8468 & 0.8860 & 0.1419 & & 0.1726 & 0.1454 & 0.1427\\
\textbf{5} & 200 & 11.63\% & & 0.8644 & 0.8980 & 0.1317 & & 0.1554 & 0.1271 & 0.1348\\
\textbf{10} & 271 & 15.76\% & & 0.8855 & 0.9145 & 0.1166 & & 0.1355 & 0.1092 & 0.1247\\\hline
\midrule
\multicolumn{11}{c}{ {\bf 2-encoder} }\\
\midrule
\textbf{1} & 209 & 12.16\% & & 0.8164 & 0.8725 & 0.1863 & & 0.1849 & 0.1422 & 0.1734\\
\textbf{3} & 357 & 20.77\% & & 0.8762 & 0.9244 & 0.1484 & & 0.1321 & 0.0838 & 0.1468\\
\textbf{5} & 422 & 24.55\% & & 0.8921 & 0.9376 & 0.1351 & & 0.1173 & 0.0696 & 0.1366\\
\textbf{10} & 528 & 30.72\% & & 0.9142 & 0.9543 & 0.1169 & & 0.0953 & 0.0519 & 0.1204\\
\bottomrule
\end{tabular}
\caption{Quantitative results: Perfect predictions, BLEU-4, and Levenshtein distance achieved by the models}
\label{tab:results}
\end{table*}

\tabref{tab:results} reports the results we achieved in the 1-encoder (top part of \tabref{tab:results}) and the 2-encoder model (bottom part). It is important to remember that the two models have been experimented exactly on the same code review instances but that the 1-encoder model has been trained/tested on the $D_p$ dataset, featuring pairs $m_s$ $\rightarrow$ $m_r$, while the 2-encoder model deals with the $D_t$ dataset, composed by triplets $\langle$$m_s$, $r_{nl}$$\rangle$ $\rightarrow$ $c_r$. In other words, when generating $m_r$, the 2-encoder model can take advantage of the comment provided by the reviewer ($r_{nl}$) and asking the specific change transforming $m_s$ into $m_r$, while this is not the case for the 1-encoder model. 

The first thing that catches the eye from the analysis of \tabref{tab:results} are the better performance ensured by the 2-encoder model. The gap, at any level of beam size, is substantial. When only one prediction is generated (\ie $k=1$) the 1-encoder model can generate the correct code in 50 cases (2.91\% of the test set) against the 209 (12.16\%) ensured by the 2-encoder model. This is a 4$\times$ improvement. The trend is confirmed for all $k$ values, with the difference, however, becoming less strong with the increase of $k$. Indeed, when 10 candidate predictions are performed, 271 perfect predictions (15.76\%) are generated by the 1-encoder model against the 528 (30.72\%) of the 2-encoder model. While the gap in performance is still notable (+94.83\% perfect predictions for the 2-enconder model), it is less marked as compared to the lowest beam size.

The BLEU-4 scores and the normalized Levenshtein distance confirm the observed trend, with the code generated by the 2-encoder model being closer to the reference code (\ie the one manually written by the developers). One observation that can be made for the 2-encoder model is that, when generating three possible previews for the code change recommended by the reviewer ($k=3$), there is one of them requiring, on average, to only change $\sim$13\% of the code tokens to obtain the reference code (median = 9\%). 

As a next step, we qualitatively analyze (i) all 271 perfect predictions obtained by the 1-encoder model with $k=10$, and (ii) all 300 perfect predictions obtained by the 2-encoder model with $k=10$ and for which the 1-encoder model failed to generate the correct prediction.

\begin{table}
\centering
\begin{tabular}{llr} 
\toprule
\multicolumn{3}{c}{\textbf{Refactoring (93)}}\\\midrule
\textbf{Method Visibility} & & {\bf 48} \\
& {\scriptsize Modifies modifier} & 20 \\
& {\scriptsize Adds modifier} & 17 \\
\vspace{0.1cm}
& {\scriptsize Removes modifier} & 11 \\
{\bf Readability} & & {\bf 42} \\
& {\scriptsize Adds/Removes curly brackets} & 8 \\
& {\scriptsize Adds/Removes ``this'' keyword} & 6 \\
& {\scriptsize Removes unneeded variable declaration} & 5 \\
& {\scriptsize Merges two code statements} & 4 \\
& {\scriptsize Removes logging information} & 4 \\
& {\scriptsize Simplifies return statement} & 4 \\
& {\scriptsize Removes parenthesis from return statement} & 3 \\
& {\scriptsize Removes unneeded ;} & 2 \\
& {\scriptsize Removes unneeded variable cast} & 2 \\
& {\scriptsize Replaces else-if with if }& 1 \\
& {\scriptsize Replaces if-else with inline if} & 1 \\
& {\scriptsize Removes unneeded object instance} & 1 \\
\vspace{0.1cm}
& {\scriptsize Removes unneeded return statement} & 1 \\
{\bf Type} & & {\bf 3} \\
& {\scriptsize Modifies variable type} & 3 \\\midrule
\multicolumn{3}{c}{\textbf{Behavioral changes (197)}} \\\midrule
 
{\bf Code Removal} & & {\bf 124} \\
& {\scriptsize If statement} & 32 \\
& {\scriptsize Method Invocation} & 31 \\
& {\scriptsize Return Statement} & 24 \\
& {\scriptsize Variable} & 21 \\
& {\scriptsize Deletes Method Body} & 15 \\
\vspace{0.1cm}
& {\scriptsize For Loop} & 1 \\
{\bf Method Invocation} & & {\bf 31} \\
& {\scriptsize Modifies parameters in method call} & 20 \\
& {\scriptsize Modifies method invocation} & 10 \\
\vspace{0.1cm}
& {\scriptsize Replaces method call} & 1 \\
{\bf Exception Handling} & & {\bf 26} \\
& {\scriptsize Removes thrown exception} & 13 \\
& {\scriptsize Removes try- catch} & 8 \\
& {\scriptsize Removes try- finally} & 4 \\
\vspace{0.1cm}
& {\scriptsize Moves variable assignment to finally block} & 1 \\
{\bf Inheritance} & & {\bf 8} \\
& {\scriptsize Removes invocation to parent's constructor} & 3 \\
& {\scriptsize Removes Override annotation} & 3 \\
& {\scriptsize Adds call to parent’s constructor} & 1 \\
\vspace{0.1cm}
& {\scriptsize Adds modifier (final)} & 1 \\
{\bf Concurrency} & & {\bf 6} \\
\vspace{0.1cm}
& {\scriptsize Removes synchronized} & 6 \\
{\bf Bug-fixing} & & {\bf 2} \\
\vspace{0.1cm}
& {\scriptsize Modifies if condition} & 2 \\
\bottomrule
\end{tabular}
\caption{Changes in the 1-encoder's perfect predictions}
\label{tab:tax1enc}
\end{table}

\tabref{tab:tax1enc} reports a classification of the code changes performed in the 1-encoder model perfect predictions. One perfect prediction can contribute to multiple categories, since several categories of changes may be performed in a single prediction. We classified each change into two macro categories, namely \emph{Refactoring} and \emph{Behavioral changes}. The former groups code transformations that we judged as unlikely of resulting in behavioral changes, while the latter should impact the code behavior. Within each macro category, a further categorization is performed to help understanding the types of code transformations learned by the model. 

In the \emph{refactoring} category, changes have been applied to the method visibility (\eg with the addition/removal/modification of {\tt public}, {\tt private}, \etc modifiers), to the type of variables (\eg change a variable declaration from {\tt HashMap $<$String, String$>$ VAR\_1 = new HashMap$<>$();} to {\tt Map $<$String, String$>$ VAR\_1 = new HashMap$<>$();}), and to improve the code readability. This sub-category features several interesting types of changes that the model recommended to simplify the code. For example, a developer submitted a method having as body {\tt TYPE\_4 $<$TYPE\_2$>$ reader = view.VAR\_1(); return reader;}. The model recommended to \emph{remove the unneeded variable declaration}, transforming the method body into {\tt return view.VAR\_1();}.  

In the \emph{behavioral changes} category, two of the implemented changes aimed at fixing bugs by modifying an {\tt if} condition. For example, in one case the model added the negation (\ie {\tt !} operator) to the if condition. Such a change was also recommended by the reviewer: ``\emph{Negation missing?}''. Note that the reviewer's comment was not available to the 1-encoder model. Also other cases in the \emph{behavioral changes} category may be related to bug-fixes but we did not have the confidence to classify them as such. For example, in the \emph{modifies parameters in method call} category, a method invocation {\tt message.substring(0, VAR\_1+1)} was changed into {\tt message.substring(0, VAR\_1)}. The reviewer's comment mentioned: ``\emph{This line will return a substring of length maxLength + 1. If the substring needs to be no longer than maxLength, then replace ``maxLength + 1'' with just maxLength}'' (VAR\_1 maps to maxLength in the abstraction map). Thus, this is likely a bug fix, assuming that the expected behavior was the one described by the reviewer. 

Due to the lack of space we do not comment on all categories in \tabref{tab:tax1enc}. However, one message that can be derived from it is that the 1-encoder model is able to learn a variety of code transformations, most of them being relatively simple in terms of code changes, but sometimes solving functional/non-functional quality issues difficult to spot.

\begin{table}
\centering
\begin{tabular}{llr} 
\toprule
\multicolumn{3}{c}{\textbf{Refactoring (9)}}\\\midrule

 {\bf Readability} & & {\bf 5} \\
 & {\scriptsize Simplify if condition} & 3 \\
 & {\scriptsize Simplify if-else statement} & 1 \\
 \vspace{0.1cm}
 & {\scriptsize Remove unneeded null check} & 1 \\
 
 {\bf Type} & & {\bf 3} \\
 \vspace{0.1cm}
 & {\scriptsize Remove type info from collection} & 3 \\
 
 {\bf Variable} & & {\bf 1} \\
 \vspace{0.1cm}
 & {\scriptsize Add modifier} & 1 \\\midrule

 \multicolumn{3}{c}{\textbf{Behavioral changes (23)}} \\\midrule
 
 {\bf Return} & & {\bf 6} \\
 & {\scriptsize Modify return type} & 4 \\
 \vspace{0.1cm}
 & {\scriptsize Modify return value} & 2 \\
 
 {\bf Code Removal} & & {\bf 4} \\
 & {\scriptsize Code block} & 3 \\
 \vspace{0.1cm}
 & {\scriptsize Switch case} & 1 \\

 {\bf Exception Handling} & & {\bf 4} \\
 & {\scriptsize Modify thrown exceptions} & 2 \\
 & {\scriptsize Modify try-catch} & 1 \\
 \vspace{0.1cm}
 & {\scriptsize Use try-with-resource pattern} & 1 \\
 
  {\bf Concurrency} & & {\bf 3} \\
 & {\scriptsize Remove concurrency lock} & 1 \\
 & {\scriptsize Remove unnecessary sync guard} & 1 \\
 \vspace{0.1cm}
 & {\scriptsize Use shared variable instead of its copy} & 1 \\

 {\bf Inheritance} & & {\bf 3} \\
 \vspace{0.1cm}
 & {\scriptsize Modify parent's constructor call} & 3 \\
 
 {\bf Bug-fixing} & & {\bf 2} \\
 \vspace{0.1cm}
 & {\scriptsize Change value of boolean} & 2 \\

 {\bf Code Addition} & & {\bf 1} \\
 \vspace{0.1cm}
 & {\scriptsize Add missing return} & 1 \\

\bottomrule
\end{tabular}
\caption{Types of changes in the 2-encoder's perfect predictions not learned by the 1-encoder}
\label{tab:2-encoder}
\end{table}

Concerning the analysis of the 2-encoder perfect predictions, here we focus on the cases in which the 1-encoder model was not able to identify the change to perform. The complete categorization of code changes we performed is available in \cite{replication}. We present here (\tabref{tab:2-encoder}) the 20 novel categories of changes we found that were not learned by the 1-encoder model. While we anlayzed 300 instances of perfect predictions performed by the 2-encoder but not by the 1-encoder, only 32 of them (11 refactorings + 21 behavioral changes) fall into categories of changes that were completely missed by the 1-encoder. This suggests that the additional comments provided as input to the 2-encoder model, while able to substantially boost its performance (528 \emph{vs} 271 perfect predictions when $k=10$), do not allow it to learn many types of code changes missed by the 1-encoder. The manual analysis also gave us the opportunity to check the effectiveness of the heuristic we use to filter out irrelevant code comments. 

We found that 22 out of the 300 inspected comments were irrelevant for the performed code changes (\ie were false positives that should have been discarded), leading to a $\sim$93\% precision for our heuristic.

Looking at \tabref{tab:2-encoder}, we can see that several of the new types of changes are still simple code changes that, however, the model was only able to learn once the reviewer's comment recommending them was provided (\eg the reviewer recommended to ``\emph{use final}'' as modifier for a variable, and the model successfully implemented the change). Others are instead more interesting, such as cases in which the model recommended to delete entire code blocks. Looking at them, in some cases the reviewers' were recommending an extract method, that was also suggested by the model. Clearly, the model limits the recommendation to the ``source'' part of the refactoring (\ie suggests which statements to extract, but not where to put them). Also, the 2-encoder model was able to learn more complex code changes that can be useful to a reviewer to quickly get a preview of how the code would look like with her comment implemented. For example, in one case the reviewer commented ``\emph{We use Java7, so you should use the try-with-resources feature}''; the 2-encoder model was able to provide as output the code implementing such a change.

As a final analysis, we looked at 100 non-perfect predictions for each model selected in the BLEU score ranges 0-24, 25-49, 50-74, and 75-99 (25 each) to determine if the recommended code change is still meaningful while being different from the reference code. For the 1-encoder, we found five (5\%) of the non-perfect predictions to be still meaningful and semantically equivalent to the code written by developers (1 in the BLEU range 50-74 and 4 in 75-99). Six (6\%) instances were instead found for the 2-encoder model (1 in 25-49, 3 in 50-74, and 1 in 75-99) as being successful cases of reviewer's comment implementation (despite being different from the change implemented by the developer). For instance, in one of these cases the reviewer asked to use for a public method the protected or default visibility. The developer replaced the {\tt public} keyword with {\tt protected}, while the 2-encoder model just removed the {\tt public} keyword, thus using the default visibility. Overall, we can estimate an additional $\sim$5\% of performance for the experimented models on top of what reported by the perfect predictions.
\section{Threats to Validity} \label{sec:threats}

\textbf{Construct validity.} While we applied many heuristics to clean the data used in the training and testing of the NMT model, by manual inspecting our dataset we still noticed a small percentage of ``noisy'' comments (\ie reviewers' comments unlikely to trigger code changes). These are due to failure of our automated heuristics and, as such, represent a limitation of the proposed approach. 
Still, NMT models should be able to deal with such a low level of noise in the data. \revised{The good results in terms of BLEU-4 and Levenshtein distance are influenced by (i) syntactic sugar of source code, and (ii) the input code provided to the model, that is usually quite similar to the one to be produced as output. This is why we complemented these analyses with additional views on the achieved results such as the number of perfect predictions.}

\textbf{Internal validity.} Subjectiveness in the manual analyses could have potentially affected our results. To mitigate such a bias, when classifying comments as \emph{relevant} or \emph{irrelevant}, two authors independently classified each comment, and a third author was involved in a case of a conflict. Also, the qualitative analyses have been performed by two of the authors. Despite this, imprecisions are still possible.

\textbf{External validity.} We mined our datasets from both Gerrit and GitHub, considering a large set of projects (8,904). However, we only focused on Java systems, thus limiting the generalizability of our findings. Still, the approach that we experimented with can be adapted to other languages by simply replacing the abstraction component.

\section{Related Work} \label{sec:related}
We summarize the related literature focusing on DL in SE and studies related to code reviews. While DL has been used to support many different SE tasks, given the goal of our work, we only focus on techniques automating source code changes.

\subsection{Deep Learning for Automating Code Changes}
Several DL-based approaches have been proposed to automatically fix bugs~\cite{Gupta:2017,Tufano:tosem2019,Chen:2019}. 

Gupta \etal propose DeepFix to fix common errors in C programs~\cite{Gupta:2017}. DeepFix is an end-to-end solution based on a multi-layered sequence-to-sequence model that iteratively tries to locate and fix errors in a given program. In the reported  evaluation, it managed to automatically fix 27\% of the testing samples. 

Similarly, Chen \etal developed SequenceR, a model designed to automatically repair bugs spanning a single line in Java code~\cite{Chen:2019}. SequenceR has been trained on a dataset of 35,578 one-line bug fixes and, in its best configuration, it was able to perfectly predict the fix for 20\% of the testing samples.

Another step in the automatic generation of fixes has been taken by Tufano \etal \cite{Tufano:tosem2019}, who evaluated the suitability of an NMT-based approach to automatically generate patches for defective code. 
Results show that the NMT models can correctly produce candidate patches for the given defective code in $\sim$9\% and $\sim$3\% of cases (depending on the length of the method) when a single patch is generated by the model.

To generalize the usability of these tools, Tufano \etal~\cite{Tufano:icse2019} investigated the possibility of using an NMT model to learn how to automatically modify a given Java method as developers would do during a pull request (PR). In other words, they try to learn generic code changes implemented over PRs using an Encoder-Decoder RNN model feeding it with methods before and after the mined PRs. They found that the model can learn a wide variety of meaningful code transformations and, in some cases, reproduce precisely the same changes that are implemented by developers in PRs.

With a similar NMT approach, Watson \etal~\cite{Watson:icse2020} tried to address one of the main open problems of automated software testing, namely the definition of the oracle. They focused on the generation of meaningful assert statements for a given test method. Their approach,  generates syntactically and semantically correct assert statements that are comparable to ones manually written by developers in 31\% of cases.

DL models have also been used to support code completion. Karampatsis and Sutton \cite{Karampatsis:DLareBest} presented an open-vocabulary neural language model to address some of the issues raised a few years before by Hellendoorn and Devanbu \cite{Hellendoorn:fse2017} on the usage of DL models for modeling code. The proposed model is able to handle new identifier names that have not appeared in its training data.
Still with the goal of supporting code completion, Kim \etal~\cite{kim2020code} first showed how Transformer models outperform the previous sequence-based models, and then focused on how to obtain higher accuracy exposing the Transformer to the syntactic structure of the code. Results show that the proposed model surpasses previous work.

Also Svyatkovskiy \etal \cite{svyatkovskiy2020intellicode} used Transformer models to support code completion. They introduced a multi-layer generative transformer model for code named GPT-C. GPT-C is the core of IntelliCode Compose, a general-purpose code completion framework able to generate syntactically correct code sequences of arbitrary token types and in multiple programming languages.

Alon \etal~\cite{alon2019structural} used DL to target code completion in a language-agnostic fashion. They presented a new approach based on LSTMs and Transformers that generates the target AST node-by-node, reaching state-of-the-art performance with an exact match accuracy for the top prediction of 18.04\%.

Another research problem in which DL has been applied is the migration of software across programming languages \cite{Nguyen:2014:ASE,Nguyen:2013:FSE:2}, \eg through statistical machine translation. 


Looking at our work in the context of the discussed literature, this is, to the best of our knowledge, the first attempt in automating code review activities from both the perspective of the contributor and of the reviewer. Form the ``technical'' point of view, the closest work to ours is the one by Tufano \etal~\cite{Tufano:icse2019} about learning code changes implemented in PRs. Indeed, their model architecture is similar to the one we use in the 1-encoder scenario (we use Transformers) and we also inherit from them the abstraction procedure. However, we target a different problem that required (i) the collection of a new dataset, with many filtering strategies put in place to avoid noise during the training process; (ii) the definition of a novel architecture exploiting two encoders to process the reviewer comment and the submitted code as input. 

\subsection{Code Review}
Several studies have focused on code reviews, investigating its impact on the code quality \cite{Kemerer:tse2009,morales2015saner,McIntosh:msr2014,Bavota:icsme2015}. Kemerer and Paulk \cite{Kemerer:tse2009}, McIntosh \etal \cite{McIntosh:msr2014}, and Bavota and Russo \cite{Bavota:icsme2015} agree in reporting the lower likelihood of introducing bugs in reviewed code as compared to non-reviewed code. Also, Morales \etal \cite{morales2015saner} confirm the higher code quality ensured by code review. 

Other authors studied how the code review process is carried out in industry and in open source communities \cite{Rigby:icse2008,Rigby:icse2011,Bacchelli:icse2013,Bosu:2013,Rigby:fse2013,BellerEtAl:2014,Czerwonka:icse2015,Kononenko:icse2016,Bosu:tse2017}. Due to space constraints, we do not discuss these works in details but we only summarize some of the findings relevant for our work. Identifying defects has been confirmed in different studies \cite{Rigby:icse2008,Bacchelli:icse2013,BellerEtAl:2014,Bosu:tse2017} as the main reason for performing code review, thus highlighting the importance of automating this activity. These studies also provided evidence of the substantial time invested in code review activities \cite{Bosu:2013,Rigby:fse2013,Czerwonka:icse2015}, one of the motivations for working on code review automation.

Researchers also analyzed the factors influencing the acceptance of the code changes submitted for review \cite{WeiBgerber:2008,Baysal:wcre2013,BosuCarver:2014A,BosuCarver:2014B}. Finally, a recent work proposed a tool, named ClusterChanges, to help developers during the code review process \cite{Barnett:icse2015}. This is the work more related to ours. However, while the overall goal is the same (\ie helping developers during code review) ClusterChanges automatically decomposes changesets submitted for a review into cohesive, smaller changes, while our goal tries to automate specific code review steps.
\section{Conclusions} \label{sec:conclusion}
We experimented with DL techniques in the context of automating two code review activities: (i) recommending to the contributor code changes to implement as reviewers would do \emph{before} submitting the code for review; and (ii) providing the reviewer with the code implementing a comment she has on a submitted code. 

This required the definition of two different transformer-based architectures.

The achieved results were promising, with the models able to generate meaningful recommendations in up to 16\% (first scenario) and 31\% (second scenario) of cases. \revised{Still, vast improvements are needed to make such models suitable to be used by developers.} 

In future, we plan to explore different DL architectures and increase the amount of data available for training our models. The latter could be the key for learning a larger variety of code changes.

\section{Data Availability}
We release the code and datasets used in our study \cite{replication}. 

\section*{Acknowledgment}
This project has received funding from the European Research Council (ERC) under the European Union's Horizon 2020 research and innovation programme (grant agreement No. 851720). Poshyvanyk was supported in part by the NSF CCF-1955853 and CCF-2007246 grants. Any opinions, findings, and conclusions expressed herein are the authors' and do not necessarily reflect those of the sponsors. 

\bibliography{main}
\bibliographystyle{IEEEtran}

\end{document}